\documentstyle[namedreferences,NATO,epsf]{CRCKAPB} 

\def\etal{{et\thinspace al.}\ }

\newcommand{\Teff}{\hbox{$T_{\rm eff}$}}

\newcommand{\ion}[2]{\mbox{#1\,{\small #2}}}
\def\re{RE\,0503-289}

\begin{opening}
\title{Metal abundances in PG1159 stars from 
\protect\\
Chandra and FUSE spectroscopy}
\author{K.\,WERNER$^1$}
\author{J.L.\,DEETJEN$^1$}
\author{S.\,DREIZLER$^1$} 
\author{T.\,RAUCH$^{2,1}$}
\institute{$^1$Universit\"at T\"ubingen, Germany\vspace{-2mm}}
\institute{$^2$Dr. Remeis-Sternwarte, Universit\"at Erlangen-N\"urnberg}
\author{M.A. BARSTOW}
\institute{University of Leicester, UK}
\author{J.W. KRUK}
\institute{Johns Hopkins University, USA}
\end{opening}
\begin{document}

\begin{abstract}
We investigate FUSE spectra of three PG1159 stars and do not find any evidence
for iron lines. From a comparison with NLTE models we conclude a deficiency of
1--1.5\,dex. We speculate that iron was transformed into heavier elements. A
soft X-ray Chandra spectrum of the unique H- and He-deficient star H1504+65 is
analyzed. We find high neon and magnesium abundances and confirm that H1504+65
is the bare core of either a C-O or a O-Ne-Mg white dwarf.
\end{abstract}

\section{Iron deficiency in PG1159 stars}

The origin of hot hydrogen-deficient post-AGB stars (spectral types [WC] and
PG1159) is supposed to be a late He-shell flash. Detailed summaries on their
spectroscopic characteristics and quantitative analyses, and relevant
evolutionary calculations can be found e.g.\ in Werner (2001) and Herwig
(2001), respectively.

In the last white dwarf workshop, held in Delaware two years ago, we presented
for the first time clear evidence for iron deficiency in a PG1159 star (Miksa
\etal 2001). Because PG1159 central stars are very hot (\Teff\ $>$75\,000\,K),
the metals are highly ionized. The dominant ionization stage of iron in the
line formation region is \ion{Fe}{VII} and most of its lines are located in the
FUV spectral region. Contrary to our expectation we were not able to detect any
\ion{Fe}{VII} line in the FUSE FUV spectrum of the pulsating central star of
K1-16, which means that iron is deficient by at least one dex. This has initiated
an effort to utilize archival IUE and HST spectra as well as new FUSE spectra
to look for iron in other PG1159 stars (Miksa \etal 2002). It turned out that
IUE and HST spectra do show a hint of iron deficiency in some other objects,
but the data quality is not sufficient for reliable quantitative analyses. The same
holds for a number of FUSE spectra from PG1159 stars. However, in two more
cases, where FUSE spectra with sufficiently high S/N were obtained, we were
successful in proving that
iron is deficient. The first one is the central star of
NGC\,7094, which is classified as a hybrid-PG1159 star, because it exhibits H
Balmer lines. No iron lines were detected and we concluded that iron is
depleted by at least 1.5 dex (Miksa \etal 2002). The next case is the central
star of Abell~78, which is classified as one of the rare [WC]--PG1159 transition
objects. For Abell~78 we find an iron deficiency of 1.5 dex from the lack of
iron lines as well (Werner \etal 2002a).

What is the origin of the iron deficiency? We think that iron has been
transformed to heavier elements. The high C and O abundances in PG1159 stars
result from envelope mixing caused by a late He-shell flash. This event also
modifies the near-solar abundance ratios of iron-peak elements in the envelope
by dredging up matter in which s-process elements were built-up by n-capture on
$^{56}$Fe seeds during the AGB phase. This scenario can and will be tested by
analyzing the resulting Fe/Ni abundance ratio, because it is significantly
changed in the intershell region in favor of Ni by the conversion of $^{56}$Fe
into $^{60}$Ni. More quantitative results from nucleosynthesis calculations in
appropriate stellar models have been presented recently (Herwig \etal 2002) and
inclusion of nuclear networks in evolutionary model sequences will become
available in the near future.

Interestingly, Asplund \etal (1999) have found that in Sakurai's object, which
is thought to undergo a late He-shell flash, iron is reduced to 0.1 solar and
Fe/Ni$\approx$3. This and other s-process signatures should also be exhibited
by Wolf-Rayet type central stars and PG1159 stars. And in fact, latest results
confirm that iron deficiency among H-deficient post-AGB stars is not restricted
to PG1159 stars. It appears that three Wolf-Rayet type central stars are iron
deficient, too. Gr\"afener \etal (2002) report a low iron abundance in SMP\,61,
an early type [WC5] central star in the LMC. Its abundance is at least 0.7\,dex
below the LMC metallicity. Crowther \etal (2002) find evidence for an iron
underabundance of 0.3--0.7\,dex in the Galactic [WC] type central stars stars
NGC~40 ([WC8]) and BD+30$^\circ$3639 ([WC9]). In this context it is also
interesting to remark that the hot DO white dwarf \re\ is iron deficient and
rich in nickel (Barstow \etal 2000). If this is related to the iron deficiency in PG1159 stars is an
open question.

\begin{figure}[ht]
\epsfxsize=\textwidth
\epsffile{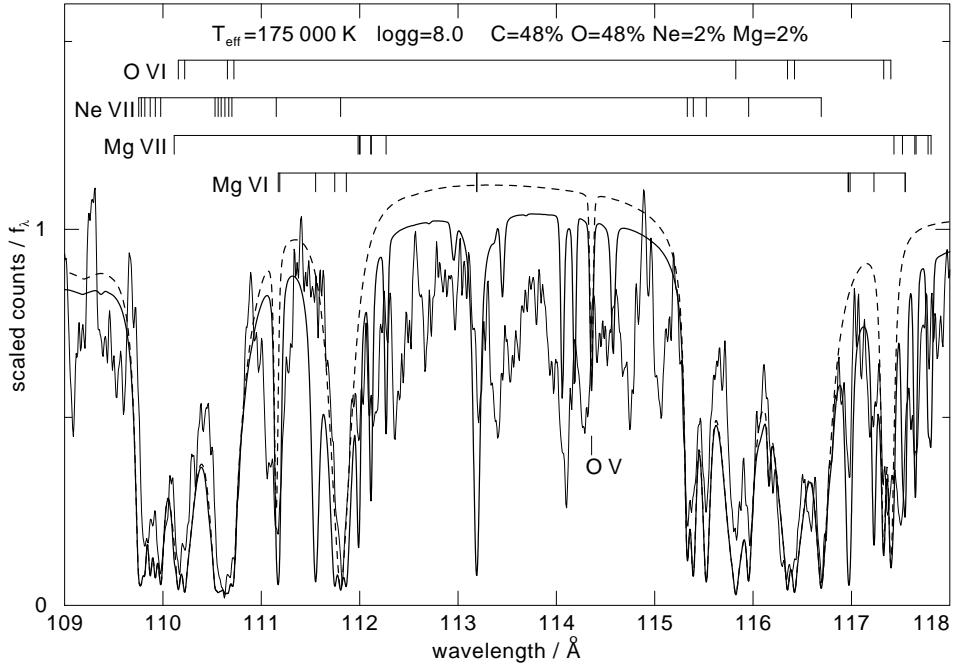}
\caption{Detail from the Chandra count spectrum of H1504+65 (thin line).
Relative fluxes of two models are overplotted, one including Mg (solid line)
and one without Mg (dashed line). Observation and model spectra are smoothed
with Gaussians (FWHM 0.02\AA\ and 0.03\AA, respectively). }
\end{figure}

\section{Chandra spectroscopy of H1504+65}

H1504+65 is the most extreme PG1159 star, not only because of its
extraordinarily high effective temperature, which is close to 200\,000\,K. We
have shown that H1504+65 is not only hydrogen-deficient but also
helium-deficient. From optical spectra it was concluded that the atmosphere is
primarily composed of carbon and oxygen, by equal amounts. Neon lines were
detected in soft X-ray spectra taken with the EUVE satellite and in an
optical-UV Keck spectrum, and a neon abundance of 2--5\% (mass fraction) was
derived (Werner \& Wolff 1999).

H1504+65 was observed with Chandra on Sep.\ 27, 2000, with an integration time
of 7 hours. Flux was detected in the range 60\AA--160\AA\ and the spectral
resolution is about 0.1\AA. The spectrum displays astonishing details
(Fig.\,1). At first sight the spectrum appears to be rather noisy, but
comparison with model spectra suggests that most absorption features in the
observation are probably real. Mg lines are discovered for the first time in
H1504+65 (see Werner \etal (2002b) for more details). Exploratory models
including opacities of heavy metals (Ca--Ni) were computed. This introduces a
very large number of additional lines, however, identifying individual heavy
metal lines will be problematic or even impossible, because the vast majority
of lines in the available lists have uncertain wavelength positions.

The Mg abundance is of the order 2\%, similar to the Ne abundance determined
earlier. This allows two alternative interpretations. Either we see a naked C-O
white dwarf with 3$\alpha$ processed matter, which is, according to
evolutionary models, dominated by C and O with an admixture of $^{22}$Ne and
$^{25}$Mg of the order 1\%. Or we see a naked O-Ne-Mg white dwarf which has
undergone carbon burning. A comparison with the respective evolutionary models
(Ritossa et al.\ 1999) suggests that we could look onto such a naked remnant
where $^{20}$Ne and a mixture of several Mg isotopes
($^{24}$Mg,$^{25}$Mg,$^{26}$Mg) should be present. We have already speculated
earlier, that the onset of C burning and subsequent evolutionary processes
might explain the uniqueness of H1504+65. In order to confirm this idea, we
will also look for Na absorption lines, because the stellar models predict a
relatively high amount of $^{23}$Na in this case. 

To conclude, H1504+65 might
have been one of the ``heavyweight'' intermediate-mass stars (9\,M$_{\odot}\le
M \le $ 11\,M$_{\odot}$) which form white dwarfs with electron-degenerate
O-Ne-Mg cores resulting from carbon burning. This speculation is corroborated
by the fact that H1504+65 is one of the most massive PG1159 stars known
(0.86$\pm$0.15\,M$_\odot$).



\begin{thebibliography}{}
\bibitem[]{a} Asplund, M., Lambert, D.L., Kipper, T., Pollacco, D., \& Shetrone, M.D.
           1999, A\&A, 343, 507
\bibitem[]{x} Barstow, M.A., Dreizler, S., Holberg, J.B., \etal 2000, MNRAS, 314, 109
\bibitem[]{b} Crowther, P.A., Abbott, J.B., Hillier, D.J., \& De Marco, O. 2002, in
        Planetary Nebulae, IAU Symp.\ 209, eds.\ M.\,Dopita et al., in press
\bibitem[]{c} Gr\"afener, G., Hamann, W.-R., \& Pe{\~ n}a, M. 2002, in
        Planetary Nebulae, IAU Symp.\ 209, eds.\ M.\,Dopita et al., in press
\bibitem[]{d} Herwig, F. 2001, in Low Mass Wolf-Rayet Stars: Origin and Evolution,
        ed.\ R.\,Waters, A.\,Zijlstra, T.\,Bl\"ocker, Ap\&SS, 275, 15
\bibitem[]{e} Herwig, F., Lugaro, M., \& Werner, K. 2002, in
        Planetary Nebulae, IAU Symp.\ 209, eds.\ M.\,Dopita et al., in press
\bibitem[]{f} Miksa, S., Deetjen, J.L., Dreizler, S., \etal 2001 in White
	Dwarfs, ed.\ J.L.\,Provencal, H.L.\,Shipman, J.\,MacDonald,
	S.\,Goodchild, ASP Conf. Ser., 226, 60
\bibitem[]{g} Miksa, S., Deetjen, J.L., Dreizler, S., Kruk, J., Rauch, T.,
	\& Werner, K. 2002, A\&A, 389, 953
\bibitem[]{h} Ritossa, C., Garcia-Berro, E., \& Iben, I. Jr. 1999, ApJ, 515, 381
\bibitem[]{j} Werner, K. 2001, in Low Mass Wolf-Rayet Stars: Origin and Evolution,
        ed.\ R.\,Waters, A.\,Zijlstra, T.\,Bl\"ocker, Ap\&SS, 275, 27
\bibitem[]{k} Werner, K., \& Wolff, B. 1999, A\&A, 347, L9
\bibitem[]{l} Werner, K., Dreizler, S., Koesterke, L., \& Kruk, J.W. 2002a, in 
	Planetary Nebulae, IAU Symp.\ 209, eds.\ M.\,Dopita et al., in press
\bibitem[]{m} Werner, K., Rauch, T., Barstow, M.A., \& Kruk, J.W. 2002b, in Exotic
	Stars as Challenges to Evolution, ed.\ C.A.\,Tout, W.\,Van Hamme, ASP
	Conf. Ser., in press
\end{thebibliography}
\end{document}